\documentclass[preprint,showpacs,preprintnumbers,amsmath,amssymb]{revtex4}

\usepackage{graphicx}
\usepackage{dcolumn}
\usepackage{bm}

\begin{document}



\title{Semiconductor saturable absorbers for ultrafast THz signals}

\author{Matthias C. Hoffmann$^1$ and Dmitry Turchinovich$^2$}


\affiliation{$^1$Max Planck Research Department for Structural Dynamics, University of Hamburg, CFEL, 22607 Hamburg, Germany \\
$^2$DTU Fotonik - Department of Photonics Engineering, Technical
University of Denmark, DK-2800 Kgs.~Lyngby, Denmark}%
\date{\today}

\begin{abstract}
We demonstrate saturable absorber behavior of n-type
semiconductors GaAs, GaP and Ge in THz frequency range at room
temperature using nonlinear THz spectroscopy. The saturation
mechanism is based on a decrease in electron conductivity of
semiconductors at high electron momentum states, due to conduction
band nonparabolicity and scattering into satellite valleys in
strong THz fields. Saturable absorber parameters, such as linear
and non-saturable transmission, and saturation fluence, are
extracted by fits to a classic saturable absorber model. Further,
we observe THz pulse shortening, and an increase of the group
refractive index of the samples at higher THz pulse peak fields.
\end{abstract}


\maketitle Semiconductor saturable absorbers and saturable
absorber mirrors (SESAMs) are routinely used for ultrafast laser
modelocking and ultrafast signal control \cite{KELLER_SESAM_1}.
Saturable absorbers operating in the visible and infrared
wavelength ranges rely on bleaching of two-level electronic
systems, usually realized by an interband transition in
semiconductor quantum wells or quantum dots \cite{Lagatsky_SESAM}.
Clearly, such quantum-confined semiconductor systems will not be
suitable for applications in the far-infrared (THz) range, where
the photon energy is much smaller than the bandgap energy of most
semiconductors, and where thermal population of such closely
spaced electronic levels would dominate. The main loss factor in
doped semiconductors in the THz frequency range is free-carrier
absorption. The THz signal attenuation is roughly proportional to
the conductivity of the material $\sigma = \mid e \mid [\mu_e n_e
+ \mu_h n_h]$, where $e$ is elementary charge, and $\mu_{e,h}$ and
$n_{e,h}$ are the mobilities and concentrations of electrons and
holes, respectively \cite{schall_jepsen, Lui_Hegmann_APL}.

In this Letter we demonstrate saturable absorbers for the THz
frequency range, based on n-type bulk semiconductors, where the
carrier mobility is modulated by nonlinear electron transport
caused by the THz electric field, thus affecting the conductivity
of the sample. Normally, the application of external electric
fields leads to acceleration of carriers in the lowest-energy
valley of the conduction band of an n-type semiconductor. At
high-momentum states the valley nonparabolicity becomes
pronounced, which leads to an increase in the effective mass and
thus to a reduction of the mobility $\mu_e$, consequently leading
to a reduced dielectric loss in the THz range
\cite{Matze_transp3}. At high enough electric fields, intervalley
scattering is possible \cite{conwell1967,constant1985}, leading to
electron transfer into the satellite valleys with reduced
curvature. This again results in a smaller electron mobility
compared to that of the conduction band minimum, and consequently
lower THz dielectric loss. Detailed studies of time-resolved
high-THz-field transport in bulk semiconductors were recently
published in Refs. \cite{Matze_transp1, Matze_transp2,
 razzari2009}.

\begin{figure} [h]
\centering
   \includegraphics[width=6.0cm]{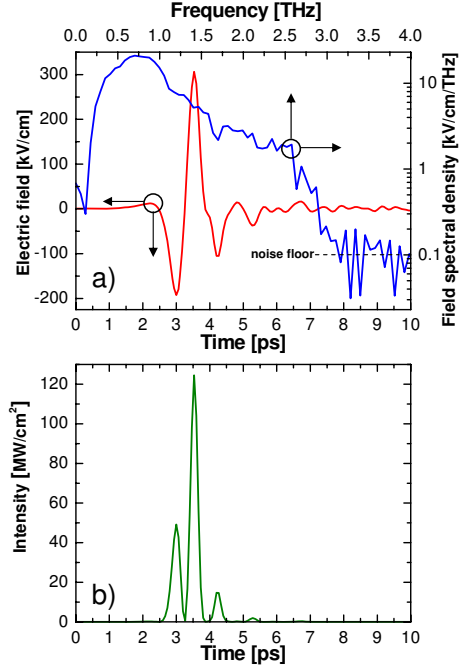}
  \caption
  {\label{fig0} (a) Time-resolved electric field of a THz pulse with a fluence of 50  $\mu$J/cm$^{2}$, and a peak electric field of
300 kV/cm. Corresponding amplitude spectrum with noise floor
indicated. (b) Instantaneous THz intensity at the sample position
calculated from the square of the measured electric field.}
   \end{figure}
In our experiment, we generated high-power single-cycle THz pulses
by tilted pulse-front optical rectification in a lithium niobate
crystal of 800-nm, 80-fs laser pulses provided by a 1-kHz
repetition rate Ti:Sapphire amplifier \cite{Matze_LNB}. The THz
pulses were collimated and then refocused onto a sample point
using a set of off-axis paraboloidal mirrors. After propagation
through the sample point, the THz pulses were guided to a 0.5-mm
thick undoped $<$110$>$-oriented ZnTe crystal for detection using
standard free-space electro-optic sampling (FEOS)
\cite{FEOS_Zhang}. A pair of wire-grid polarizers was introduced
into the THz beam path before the sample point, which allowed us
to controllably attenuate the THz signal by adjusting the angle
between the polarizer axes. The maximum THz pulse energy was
measured to be 1.5 $\mu$J using a calibrated pyroelectric
detector, in the THz beam waist of 1.0 mm (intensity, 1/$e^2$) at
the sample location. This is equivalent to a maximal THz fluence
of $F_{max}= 50$ $\mu$J/cm$^{2}$, corresponding to a peak electric
field of 300 kV/cm. The peak field was determined by calibrating
the squared time-domain electro-optic signal with the known value
of pulse fluence. Fig. \ref{fig0} shows the temporal dependency of
electric field as measured by FEOS, the amplitude frequency
spectrum, and corresponding instantaneous intensity at the
position of the sample of the strongest THz signal used in our
experiments. The samples under study were n-type semiconductors:
GaAs with a carrier concentration of $8\times10^{15} \,
\textrm{cm}^{-3}$ and a thickness of 0.4 mm; GaP with a carrier
concentration of $10^{16} \, \textrm{cm}^{-3}$ and thickness of
0.3 mm; and Ge with a carrier concentration of $10^{14} \, \textrm
{cm}^{-3}$ and a thickness of 6 mm. The measurements were carried
out at room temperature.
\begin{figure} [h]
\centering
   \includegraphics[width=8.0cm]{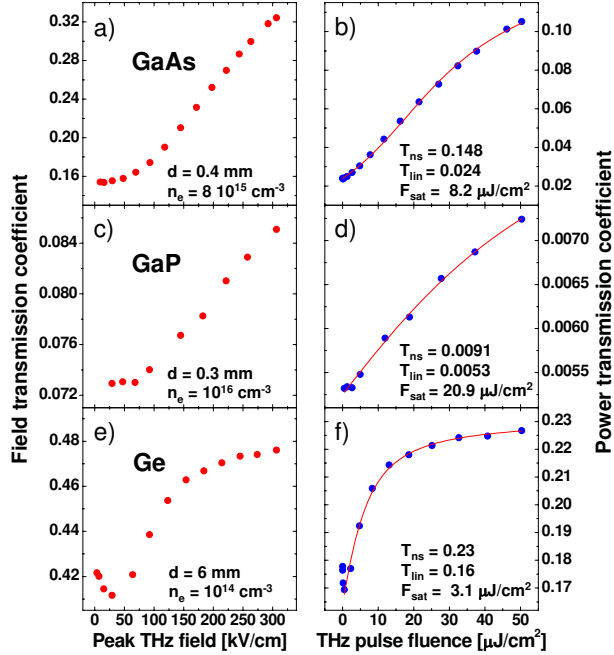}
  \caption
  {\label{fig1} Symbols: field transmission coefficient as a function of peak THz pulse field, and power transmission coefficient as a function
  of THz pulse fluence for GaAs (a-b), GaP (c-d), and Ge (e-f). Solid lines - saturable transmission function fit to the measured power transmission
  coefficients. See text for details.}
   \end{figure}
Fig. \ref{fig1} shows the field and power transmission
coefficients as a function of THz pump field and fluence
respectively in GaAs, GaP and Ge. The field and power transmission
coefficients were obtained by integrating either the modulus or
the square, respectively, of the transmitted THz fields
transmitted through the sample, and dividing them by reference
values recorded without the sample in the beam path. In all our
samples we observed increased transmission at higher pump
fluences. In particular, we observed a nearly five-fold increase
in power transmission coefficient for GaAs sample in the THz pulse
fluence range used in our experiments. The solid lines in Figs.
\ref{fig2}(b,d,f) are fits using a saturable power transmission
function, defined after Ref. \cite{Keller_SESAM} as
\begin{equation}\label{SatTransFormula}
    T(F_{p})=T_{ns} \frac{ln[1 + T_{lin}/T_{ns}(e^{F_{p}/F_{sat}}-1)]}
    {F_{p}/F_{sat}},
\end{equation}
where $T_{lin}$ and $T_{ns}$ are linear and non-saturable power
transmission coefficients, $F_{p}$ is the pump fluence, and
$F_{sat}$ is the saturation fluence. Using fits with Eq.
\ref{SatTransFormula} we were able to extract the saturable
absorber parameters for our semiconductor samples. These
parameters, namely linear and non-saturable transmission, and
saturation fluence are indicated in Fig. \ref{fig1}. In
particular, the saturation fluence $F_{sat}$ was found to be 8.2
$\mu$J/cm$^{2}$ for GaAs, 20.9 $\mu$J/cm$^{2}$ for GaP, and 3.1
$\mu$J/cm$^{2}$ for Ge. Interestingly, at the lowest pump fluences
the transmission function for Ge actually decreases slightly, thus
demonstrating the opposite, optical limiting behavior. This can be
possibly explained by the fact that in Ge the high-mobility
$\Gamma$-valley is not initially populated, and thus initial
sample conductivity is lower, as discussed in Ref.
\cite{Matze_transp3}.
\begin{figure} [h]
\centering
   \includegraphics[width=5.0cm]{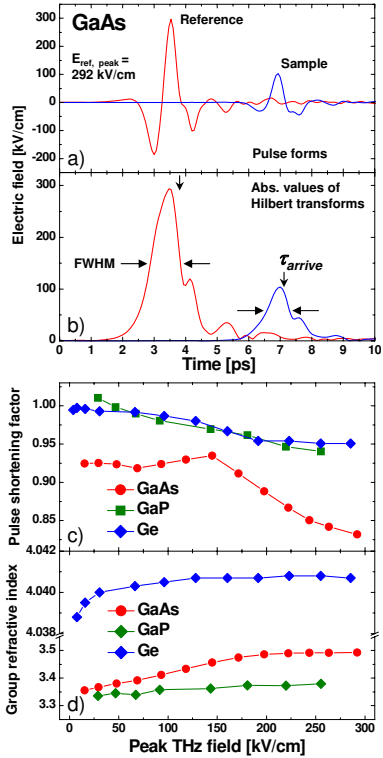}
  \caption
  {\label{fig2} (a) Shape of the THz pulses before (reference) and after (sample) propagating through the 0.4-mm thick GaAs
  sample. This reference pulse had peak electric field strength of
  292 kV/cm. (b) Modulus of Hilbert transforms of these THz pulses. Horizontal arrows indicate FWHMs of the pulses, and vertical arrows indicate the mean-weighted
  arrival times of the pulses. (c) Pulse shortening factors and (d) group refractive indices of GaAs, GaP, and Ge
  samples as functions of peak electric field of incident THz pulse. See text for details.}
   \end{figure}
In all our samples we observed shortening of the THz pulses, as
they propagated through the samples, with increase in peak
electric field of the THz pulse. Such pulse-shortening behavior is
also characteristic for saturable absorbers in the optical range.
In order to quantify the shortening of \emph{single-cycle} THz
pulses used in our experiment, we calculated a pulse shortening
factor defined as the ratio of full width at half maximum (FWHM)
of modulus of Hilbert transforms of sample and reference pulses,
as illustrated in Fig. \ref{fig2}(a,b). The dependency of the
pulse shortening factor on the peak electric field of the incident
THz pulse is shown in Fig. \ref{fig2}(c). For GaP and Ge the
maximum relative pulse shortening (i.e. the difference between
largest and smallest shortening factors) reaches approximately
5{\%}, and its dependency on THz peak field is more or less
linear. For GaAs, however, this dependency appears to have a
threshold at THz peak field of around 150 kV/cm and, and the
relative pulse shortening reaches the value of approximately
10{\%}.

Saturable absorption and pulse shortening are accompanied by an
increase in group refractive index in all three samples, as shown
in Fig. \ref{fig2}(d). The group index $n_g$ at various pump
intensities was calculated using the difference of arrival times
$\Delta \tau$ between sample and reference pulses and taking into
account the sample thickness $d$ by using the relation $n_g =
\Delta \tau c / d + 1$. The arrival times were obtained from
mean-weighted maxima of the modulus of THz pulse Hilbert
transforms. We observe a clear saturating growth trend in the
group refractive indices of all our samples with increasing THz
fields. The most dramatic index growth is observed in GaAs, and it
reaches $\Delta n_g = 0.14$, whereas for Ge it is only $\Delta n_g
= 0.002$ in the full THz pulse peak field range available in our
experiments. We note that very similar dependencies were obtained
if the maxima of THz pulses, and not of the modula of their
Hilbert transforms were analyzed.

The observed increase in group index with stronger THz pump is
indeed anticipated, and two contributions to this effect are
considered. The first contribution is the change in reflectivity
at the sample \emph{surface} \cite{schall_jepsen}. A decrease in
the sample absorption, resulting from the strong THz excitation,
leads to a positive phase shift of THz pulse at the interface
between the air and absorbing sample. This causes a temporal delay
of the transmitted THz pulse and consequently a larger group
refractive index. The other contribution to a group index growth
with stronger pump stems from the propagation through the
\emph{bulk} of the sample with the phase index dispersion in the
frequency range below resonance (in this case - plasma resonance).
Larger effective mass $m^*$, acquired by the carriers at
high-momentum states as a result of intense THz excitation, will
lead to a decrease in plasma frequency $\omega_p =  (n_e e^2 /
\epsilon_0 m^*)^{1/2}$, thus making the frequency dependency of
phase index $n_{\phi}(\omega)$ more steep. This would result in
the higher value of a group index $n_g = n_{\phi} + \omega \, d
n_{\phi} / d \omega$ for the stronger excited samples. The same
effect will arise from the effective reduction of the density of
highly mobile carriers $n_e$ at the bottom of the $\Gamma$-valley.
Both interface and bulk contributions to group index growth
obviously depend on the strength of the THz excitation and will
thus saturate with the overall THz transmission, which is in
accordance with our observations. We note that a similar increase
in the pulse delay and thus in the group index with stronger
optical pump was also observed in saturable absorbers for the
optical range, as described in Ref. \cite{mork}.

We further note that n-type GaP is often used as a broadband FEOS
detector, where parasitic etalon reflections of the THz pulses
within the crystal are conveniently suppressed by the free-carrier
absorption. As we have shown here, in the case of high THz field
strengths the results of THz spectroscopy will be distorted due to
n-GaP nonlinearities. Therefore, for etalon suppression in
high-field experiments it is advisable to use combined
active-inactive crystals \cite{dmt_combined_xtals} instead of
doped crystals.

In conclusion, we have observed ultrafast saturable absorption in
n-type semiconductors GaAs, GaP, and Ge in the THz range at room
temperature. We were able to extract linear and non-saturable
transmission coefficients, as well as the saturation fluence using
a standard saturable absorber model developed for the optical
frequency range. Our data shows THz pulse shortening, and growth
of group refractive index of our samples with increase in THz
excitation, correlated with the saturating transmission through
the samples. Although the origin of the absorption mechanism in
the THz and in the optical frequency ranges is completely
different, we note that the saturation fluences observed here are
within the same order of magnitude (i.e. few to tens of
$\mu$J/cm$^{2}$), as the values reported for SESAMs in the optical
range, such as e.g. quantum dot SESAMs \cite{Lagatsky_SESAM,
dmt_QDs_satabs}. Optimized THz SESAMs could potentially be used
with the THz sources like quantum cascade lasers (QCLs), leading
for example to external-cavity mode-locked QCLs delivering
high-energy ultrashort THz pulses.

We are grateful to Danish Advanced Technology Foundation (HTF) for
financial support; and to A. Cavalleri (Univ. Hamburg), K. Yvind,
J. M{\o}rk, and J. M. Hvam (DTU Fotonik) for valuable assistance
and discussions. Correspondence should be addressed to M.C.H
(matthias.c.hoffmann@desy.de) or D.T. (dmtu@fotonik.dtu.dk).

\pagebreak


\pagebreak

\textbf{Figure captions}\\
\\
\emph{Caption to Fig. 1}\\
(a) Time-resolved electric field of a THz pulse with a fluence of
50  $\mu$J/cm$^{2}$, and a peak electric field of 300 kV/cm.
Corresponding amplitude spectrum with noise floor indicated. (b)
Instantaneous THz intensity at the sample position calculated from
the square of the measured electric field.
\\
\emph{Caption to Fig. 2}\\
Symbols: field transmission coefficient as a function of peak THz
pulse field, and power transmission coefficient as a function of
THz pulse fluence for GaAs (a-b), GaP (c-d), and Ge (e-f). Solid
lines - saturable transmission function fit to the measured power
transmission coefficients. See text for details.
\\
\emph{Caption to Fig. 3}\\
(a) Shape of the THz pulses before (reference) and after (sample)
propagating through the 0.4-mm thick GaAs sample. This reference
pulse had peak electric field strength of 292 kV/cm. (b) Modulus
of Hilbert transforms of these THz pulses. Horizontal arrows
indicate FWHMs of the pulses, and vertical arrows indicate the
mean-weighted arrival times of the pulses. (c) Pulse shortening
factors and (d) group refractive indices of GaAs, GaP, and Ge
samples as functions of peak electric field of incident THz pulse.
See text for details.

\end{document}